\documentclass[%
preprint,
showpacs,
preprintnumbers,
nobibnotes,
amsmath,amssymb,
aps,
prb,
longbibliography,
]{revtex4-1}

\usepackage{graphicx}
\usepackage{dcolumn}
\usepackage{bm}
\usepackage{hyperref}
\usepackage{color}
\usepackage{float}

\begin{document}

\title{
  Ab-initio studies of exciton $\texorpdfstring{\boldsymbol{g}}{g}$ factors:\\
  Monolayer transition metal dichalcogenides in magnetic fields
}

\author{Thorsten Deilmann}
  \email{thorsten.deilmann@wwu.de}
  \affiliation{Institut f\"ur Festk\"orpertheorie, Westf\"alische Wilhelms-Universit\"at M\"unster, 48149 M\"unster, Germany}%
\author{Peter Kr\"uger}
  \affiliation{Institut f\"ur Festk\"orpertheorie, Westf\"alische Wilhelms-Universit\"at M\"unster, 48149 M\"unster, Germany}%
\author{Michael Rohlfing}
  \affiliation{Institut f\"ur Festk\"orpertheorie, Westf\"alische Wilhelms-Universit\"at M\"unster, 48149 M\"unster, Germany}%

\date{April 30, 2020}

\begin{abstract}
The effect of a magnetic field on the optical absorption in semiconductors
has been measured experimentally and modeled theoretically for various systems
in previous decades.
We present a new first-principles approach
to systematically determine
the response of excitons to magnetic fields, i.e. exciton $g$ factors.
By utilizing the $GW$-Bethe-Salpeter equation methodology
we show that $g$ factors extracted from the Zeeman shift of electronic bands
are strongly renormalized by many-body effects
which we trace back to the extent of the excitons in reciprocal space.
We apply our approach to
monolayers of transition metal dichalcogenides (MoS$_2$, MoSe$_2$, MoTe$_2$, WS$_2$, and WSe$_2$)
with strongly bound excitons
for which $g$ factors are weakened by about 30\%.
\end{abstract}

\maketitle

\section{Introduction}
The response of a semiconductor
to magnetic fields is intimately linked to its quantum mechanical properties.
The two main effects,
the Zeeman and the diamagnetic shift,
have been employed by many researchers
to study the electronic and optical properties of,
e.g., bulk semiconductors \cite{Wilson_1969,Yafet1963},
quantum dots \cite{PhysRevLett.82.1748,PhysRevB.59.R10421},
or recently atomically thin materials,
e.g., two-dimensional transition metal dichalcogenides (TMDCs)
\cite{Li_2014,Aivazian_2015,Srivastava_2015,Arora_2016}.
The nature of the quantum mechanical states
involved in optical processes
(excitons, i.e. electron-hole pairs) determines
its shift in the magnetic field, the so-called exciton $g$ factor.
Due to the unique character of each excitation,
$g$ factors allow for their identification and
the analysis of their properties.
E.g. interlayer excitons in bulk MoTe$_2$ have been
discoverd by the different sign of its $g$ factor \cite{MoTe2bulk}.
In TMDCs often a variety of exciton lines
with different shifts are found with
measured values of about $-4$, $-8$, or even stronger \cite{Koperski_2018}.
For higher excited Rydberg excitons $2s$, $3s$, etc.
stronger $g$ factors than for its $1s$ counterpart have been reported
\cite{Chen_2019,Goryca_2019,Liu_Rydberg_2019}.
Changing values have also been observed in
temperature and doping dependent measurements \cite{Wang-mag_2018,Liu_2020}.
At present, however, a conclusive understanding and ab-initio prediction
of exciton $g$ factors is still missing.

On a single particle level,
the fundamental theory of semiconductors in magnetic fields
has been formulated by Kohn \cite{Kohn_1959} and Roth \cite{Roth_1962} around the 1960s.
For the calculation of the magnetization,
i.e. the ${\bf k}$-integrated magnetic moment,
it has later been reformulated
on the basis of the Berry phase,
e.g. for ferromagnets
\cite{Chang_1996,Xiao_2005,Thonhauser_2005,Xiao_2010,PhysRevB.94.121114}.
Most theoretical descriptions
to evaluate exciton $g$ factors
are based on models applying ${\bf k \cdot p}$ theory \cite{Willatzen_2009,Roth_1959,PhysRevLett.96.026804,Kormanyos_2015,Koperski_2018,Arora_2018,PhysRevB.97.085153,Rybkovskiy_2017,Rostami_2015}.
Within these approaches, however, only the individual magnetic moments of \textit{single} electrons and holes are considered while 
the \textit{excitonic} many-body nature of the correlated electron-hole pair is typically ignored.

In this Letter, we present a new first-principles approach
merging the evaluation of the ${\bf k}$-resolved orbital magnetic moments
and the properties of the excitons
to calculate its $g$ factors.
These calculations employ ab-initio density functional theory (DFT) and the 
$GW$-Bethe-Salpeter equation (BSE) \cite{Rohlfing_eh,RevModPhys.74.601}.
While the explicit use of magnetic fields is challenging in a self-consistent approach,
it is certainly possible to utilize the wave functions to calculate
band- and ${\bf k}$-dependent magnetic moments from perturbation theory.
To this end, we rewrite the original approach \cite{Kohn_1959,Roth_1962}
into the form of Chang et al. \cite{Chang_1996}.
The resulting magnetic moments
take into account the full Bloch states, i.e. the calculations are beyond
a local approximation which treats 
only contributions from  small spheres around the atoms.
To evaluate exciton $g$ factors
we consider the spatial structure of the excitation
gained from the BSE.
After introducing our approach for MoSe$_2$,
we discuss the results for the five well-known TMDC monolayers
MoS$_2$, MoSe$_2$, MoTe$_2$, WS$_2$, and WSe$_2$
and compare them to experiment.
We show that 
exciton $g$ factors based on full Bloch states are enhanced
with respect to those from the local approximation 
while they are weakened by their many-body character.

A quantum mechanical system in a homogeneous magnetic field
(we use ${\bf B} = (0,0,B_z)$ as in most experiments)
is described by the effective one-particle Hamiltonian
\vspace*{-.2cm}
\begin{align}\label{eq:H}
  \hat{H}^\text{eff}& = \hat{H}_0^\text{eff} + \frac{e}{2m_e}(\hat{L}_z + g_e \hat{S}_z) B_z + \frac{e^2}{2m_e}(x^2+y^2)B_z^2 \nonumber\\
                    & =: \hat{H}_0^\text{eff} - \hat{m}_z B_z + \mathcal{O}(B_z^2) ,
\end{align}
where $\hat{H}_0^\text{eff} | \Psi_{n{\bf k}}^0 \rangle = E_{n{\bf k}} |\Psi_{n{\bf k}}^0 \rangle$ is the non-perturbed
DFT or $GW$ Hamiltonian, respectively, including spin-orbit coupling.
$\hat{L}_z$ and $\hat{S}_z$ are angular momentum and spin operator and $g_e$ is the free electron $g$ factor.
In this study we will focus on the linear Zeeman term $\hat{m} B$ (we omit the index $z$ for brevity)
and neglect the diamagnetic term which is quadratic in $B$.
The magnetic moment can be further divided into
\vspace*{-.2cm}
\begin{align}\label{eq:m}
  \hat{m} =: \hat{m}^\text{orb} + \hat{m}^\text{spin} = -\frac{e}{2m_e} \hat{L} -\frac{e g_e}{2m_e} \hat{S} .
\end{align}
In the case of an isolated hydrogen atom
these numbers correspond to the magnetic quantum number $m_l$
and the spin quantum number $m_s$.
In a periodic semiconductor its expectation value for band $n$ at $\bf k$ can be calculated by
$m_{n{\bf k}} = \langle \Psi_{n{\bf k}}^0 | \hat{m}^\text{orb} | \Psi_{n{\bf k}}^0 \rangle
+ \langle \Psi_{n{\bf k}}^0 | \hat{m}^\text{spin} | \Psi_{n{\bf k}}^0 \rangle$.

\section{Magnetic moments of Bloch states}
While the calculation of the spin part in Eq.~(\ref{eq:m}) is easy once the spinors are known,
the evaluation of the orbital part is more delicate.
The spatial dependency of the operator $\hat{L}_z = x\hat{p}_y - y\hat{p}_x$
prevents a straightforward calculation
and we will discuss two different approaches:
(i) The most simple way to tackle the problem is
a local approximation as it has been carried out earlier for the magnetization \cite{PhysRevB.72.024452,PhysRevB.94.121114}.
Within this approximation, we can easily evaluate
\begin{align}\label{eq:morb}
  m_{n{\bf k}}^\text{orb,loc} = \langle \Psi_{n{\bf k}} | x\hat{p}_y - y\hat{p}_x | \Psi_{n{\bf k}} \rangle_\text{loc}
\end{align}
in a basis of Gaussian orbitals \cite{wieferink_improved_2006} in a local sphere around each atom in the unit cell,
yielding the data of Fig.~\ref{fig:MoSe2_m}(a).
However, in regions of the Brillouin zone where the Berry curvature is large (see Supplemental Material)
this approximation for the magnetic moments clearly fails (see Fig.~\ref{fig:MoSe2_m}(a,b)).
(ii)~To account for the full Bloch states,
i.e. considering the different contributions of $\hat{L}_z$
with the lattice-periodic wave function in different unit cells,
we follow the approach proposed by Kohn \cite{Kohn_1959} and Roth \cite{Roth_1962}
\begin{align}\label{eq:morb-orig}
  \bar{m}_{n{\bf k}}^\text{orb} = -\frac{i \mu_\text{B}}{m_e} \sum_{n'\not=n}
    \left( \frac{\langle u_{n{\bf k}}|\hat{p}_x|u_{n'{\bf k}}\rangle  \langle u_{n'{\bf k}}|\hat{p}_y| u_{n{\bf k}}\rangle}{E_{n'{\bf k}} - E_{n{\bf k}}} \right. - \nonumber\\
    \left. \frac{\langle u_{n{\bf k}}|\hat{p}_y|u_{n'{\bf k}}\rangle  \langle u_{n'{\bf k}}|\hat{p}_x| u_{n{\bf k}}\rangle}{E_{n'{\bf k}} - E_{n{\bf k}}} \right) .
\end{align}
By using the commutator relation $i [\hat{H},x_j] = \frac{\hbar}{m_e} \hat{p}_{x_j}$
one can transform Eq.~(\ref{eq:morb-orig}) to Eq.~(\ref{eq:morb-new}),
which has been derived by Chang et al. \cite{Chang_1996} for the magnetic moment of a wave packet
\begin{align}\label{eq:morb-new}
  m_{n{\bf k}}^\text{orb} = \mu_\text{B} \text{Im} \left\langle \frac{\partial u_{n{\bf k}}}{\partial k_x} \middle| \hat{H}({\bf k}) - E_{n{\bf k}} \middle| \frac{\partial u_{n{\bf k}}}{\partial k_y} \right\rangle.
\end{align}
Taking special care on the nonlocal pseudopotential \cite{Pickard_2000,Kim_2018},
we find that both, Eq.~(\ref{eq:morb-orig}) and Eq.~(\ref{eq:morb-new}) lead to equivalent magnetic moments
at the $\pm$K point
if we consistently use DFT or $GW$ energies, respectively, and the corresponding wave functions \cite{PhysRevLett.63.1719}.
We note that the increased gap in $GW$ \cite{HedinGW,GdWTMDC}
generally leads to larger magnetic moments.
Due to the numerical stability (Eq.~(\ref{eq:morb-orig}) diverges for degenerated states)
we employ Eq.~(\ref{eq:morb-new}) in the following
yielding the data of Fig.~\ref{fig:MoSe2_m}(b).

\begin{figure*}[tb]
  \centering
  \includegraphics[width=\linewidth]{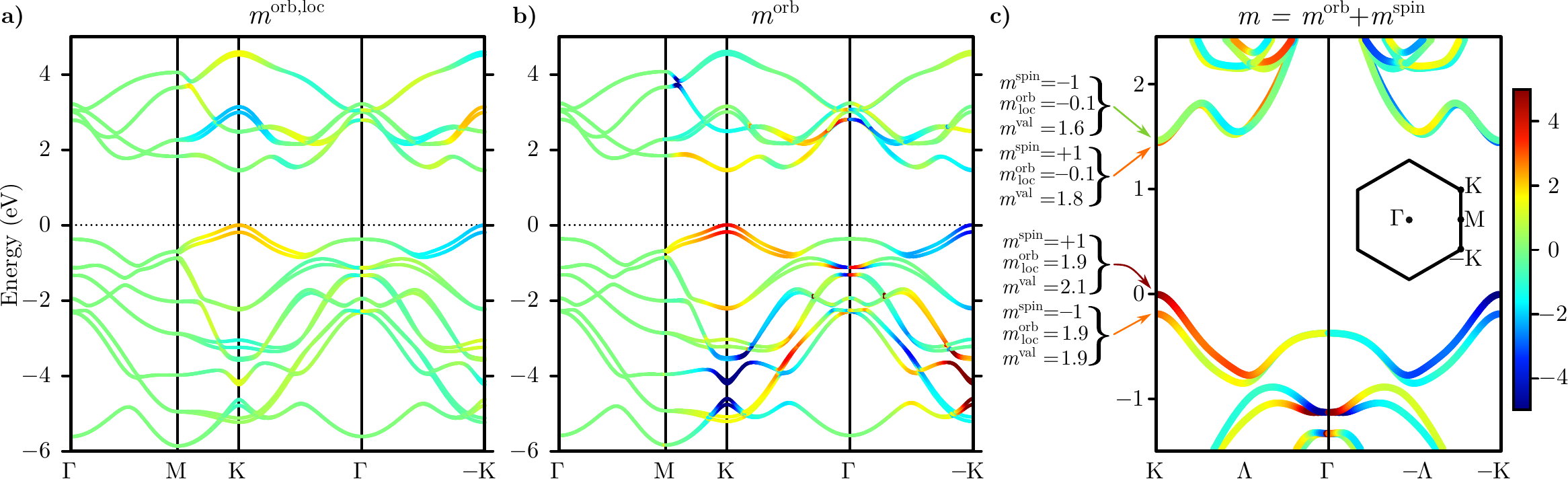}
  \caption{
    DFT band structure of MoSe$_2$.
    In a) the bands are colored according to
    the expectation values of the local orbital magnetic moments $m^\text{orb,loc}_{n{\bf k}}$ (Eq.~(\ref{eq:morb})).
    In contrast to this, in b) the orbital magnetic moment $m^\text{orb}_{n{\bf k}}$
    including the contribution from the Bloch states is shown (Eq.~(\ref{eq:morb-new})).
    The corresponding color scale is shown in units of $\mu_\text{B}$.
    In c) a zoom in along K$\Gamma -$K is shown in which the spin contributions have been added.
    The colors refer to the total magnetic moments $m_{n{\bf k}} = m^\text{orb}_{n{\bf k}} + m^\text{spin}_{n{\bf k}}$.
    On the left side we break down the different contributions of $m$ at the K point (see main text).
  }\label{fig:MoSe2_m}
\end{figure*}
\section{Magnetic moments of the bands in $\text{MoSe}_2$}
In a TMDC monolayer like e.g. MoSe$_2$ each band consists
of a superposition of different orbitals of different atoms.
Close to the K point the character of the topmost valence bands is dominated by the
Mo atoms with a contribution of more then 80\%
which stems from the $d$ orbitals,
whose major part of about 90\% is related to the spherical harmonics $Y_{2,\pm 2}$ ($d_{x^2-y^2}$ and $d_{xy}$ orbitals).
The remaining 18\% are shared by $p_x$ and $p_y$ orbitals of Se atoms.
In contrast to this, the lowest conduction band is dominated
by Mo $Y_{2,0}$ ($d_{z^2}$ orbital) with a share of about 55\%.
In Fig.~\ref{fig:MoSe2_m}a) the resulting local magnetic moment (Eq.~(\ref{eq:morb})) is shown.
Indeed, as the discussion of the special harmonics suggests,
we find a value of $m^\text{orb,loc}_{\text{VB},\text{K}} = 1.93\,\mu_\text{B}$ for both spin-orbit split
valence bands and of $m^\text{orb,loc}_{\text{CB},\text{K}} = -0.09\,\mu_\text{B}$ for the conduction bands.
Further away from the K point the moments almost vanish.
At $-$K the sign of the orbital moment is exactly reversed.
In general we find the orbital magnetic moments of MoSe$_2$ to be
$-2 \leq m^\text{orb}_\text{loc} \leq 2$ for bands close to the Fermi level,
which corresponds well to the $s$, $p$, and $d$ wave function character.
Note that, e.g. in the $\Gamma$M direction, the magnetic moments are zero
even though strong $p$ and $d$ characters are observed,
which underlines the importance of the relative phase of the contributing orbitals in the superposition.

However, this local approximation is over-simplified.
In a periodic semiconductor the correct physical states are Bloch waves.
Taking this into account \cite{Kohn_1959,Roth_1962} reveals
several important quantitative differences (Fig.~\ref{fig:MoSe2_m}b)).
While the orbital momentum at $\Gamma$ remains zero,
the situation at $\pm$K is distinctly changed.
For the valence bands we find slightly different values
close to $4$ while $m^\text{orb}$ of the conduction bands
is slightly smaller then $2\,\mu_\text{B}$ based on DFT.
The difference to the local picture is
given as the so-called valley contribution \cite{Koperski_2018}
$m^\text{val} = m^\text{orb}-m^\text{orb}_\text{loc} = 2.1$
and $1.8\,\mu_\text{B}$, respectively.
When employing $GW$ the deviations from the local approximation are even larger
and we calculate $m^\text{orb} = 5.6$ and $3.2\,\mu_\text{B}$
($m^\text{val} = 3.7$ and $3.3\,\mu_\text{B}$), respectively.
If not noted explicitly, we will stick to the DFT
results and refer to the Supplemental Material for further discussion.

For calculating the entire magnetic moment the spin part is still missing.
In Fig.~\ref{fig:MoSe2_m}c) we show a zoom-in
with the bands colored according to 
$m_{n{\bf k}} = m^\text{orb}_{n{\bf k}} + m^\text{spin}_{n{\bf k}}$.
At the $\pm$K points the valence and conduction bands
have opposite spin direction due to the spin-orbit interaction.
Hence, at K the two topmost valence bands have a magnetic moment
of about $m_{\text{VB},\text{K}} = 2.8$ and $5.0\,\mu_\text{B}$, respectively.
The same happens for the two lowest conduction bands which are close in energy
($m_{\text{CB},\text{K}} = 2.7$ and $0.5\,\mu_\text{B}$).
Also the spin part acts with a reversed sign at $-$K
so that $m_{n,\text{K}} = -m_{n,-\text{K}}$ holds \cite{Sundaram_1999}.
We observe similar but quantitatively different results for all TMDC monolayers.
We will subsequently discuss and compare the numbers
(see Tab.~\ref{tab:mg} below).

\section{Evaluation and interpretation of exciton $g$ factors}
We now deduce the effects of a small magnetic field on \textit{excitons}.
In an over-simplified picture
an exciton would be a transition from one
point ${\bf k}$ in a valence band
to another point ${\bf k}+{\bf Q}$ in a conduction band,
and its change with the magnetic field would be given
by the difference $m_{c{\bf k +Q}} - m_{v{\bf k}}$.
However, this approximation
(which means neglecting the electron-hole interaction)
is known to be unsatisfactory for most systems
and, in particular, for 2D systems with large exciton binding energies \cite{c2db}.
The state-of-the-art approach to account for two-particle excitations
is the BSE \cite{Strinati82,Rohlfing_eh,RevModPhys.74.601}
which is given in the Tamm-Dancoff approximation by
\begin{align}\label{eq:BSEq}
  (&\epsilon_{c{\bf k}+{\bf Q}} - \epsilon_{v{\bf k}}) A^{(N,{\bf Q})}_{vc{\bf k}}
  + \sum_{v'c'{\bf k}'} K_{vc{\bf k},v'c'{\bf k}'}({\bf Q}) A^{(N,{\bf Q})}_{v'c'{\bf k}'} \nonumber\\
  &= \Omega^{(N,{\bf Q})} A^{(N,{\bf Q})}_{vc{\bf k}} .
\end{align}
Here, $\Omega^{(N,{\bf Q})}$ is the energy of exciton $N$
and $A^{(N,{\bf Q})}_{vc{\bf k}}$ its amplitudes,
which contain the complete spatial structure.
Again we assume moderate magnetic fields,
i.e. that the change of electron-hole interaction $K_{vc{\bf k},v'c'{\bf k}'}({\bf Q})$ due to the field can be neglected.
Consequently, the influence of a magnetic field on 
the energy of an exciton is given by the field induced
change of the band-structure energy differences $\epsilon_{c{\bf k}+{\bf Q}} - \epsilon_{v{\bf k}}$
of all contributing transitions.
We can eventually evaluate the effective exciton $g$ factor of the exciton $N$ with momentum ${\bf Q}$ by
\begin{align}\label{eq:g}
  g^{(N,{\bf Q})} = 2 \sum_{vc{\bf k}} |A^{(N,{\bf Q})}_{vc{\bf k}}|^2 (m_{c{\bf k +Q}} - m_{v{\bf k}}) / \mu_\text{B} .
\end{align}
In experiment (effective) $g$ factors are typically defined on the basis of the
energy difference between measurements with right- and left-handed circular polarized light
$g \mu_B B := \Omega_{\sigma^+} - \Omega_{\sigma^-}$ \cite{stier_exciton_2016}.
This results in the factor $2$ in Eq.~(\ref{eq:g}).
If excitonic effects were neglected,
the $g$ factor of the transition from ($v$, ${\bf k}$) to ($c$, ${\bf k+Q}$) could be approximated by
\begin{align}\label{eq:gband}
  g^{(v{\bf k}\to c{\bf k+Q})}_\text{band} = 2 (m_{c{\bf k+Q}} - m_{v{\bf k}}) / \mu_\text{B}.
\end{align}
Eq.~(\ref{eq:g}) is a generalization of $g_\text{band}$.
Our results show that $g$ is more than 30\% smaller compared to $g_\text{band}$
due to the spatial structure of the excitons.

\begin{figure*}[t]
  \centering
  \includegraphics[width=.98\linewidth]{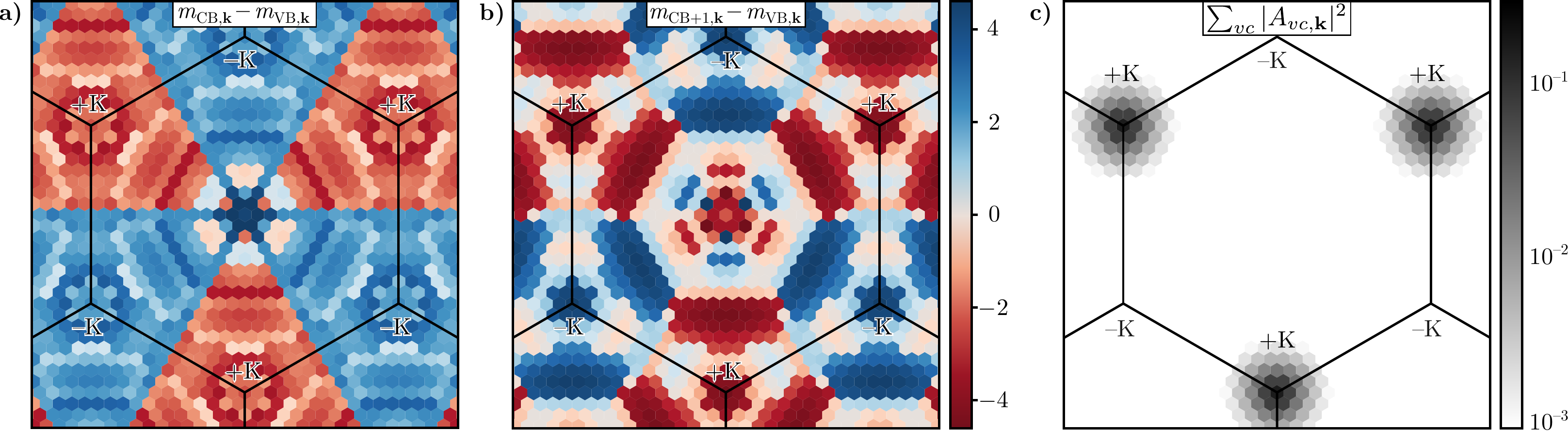}
  \caption{
    a,b) Difference of the magnetic moments of MoSe$_2$'s first and second conduction band (CB, CB+1), respectively,
    and the valence band (VB).
    The results are shown on a $(24\times 24)$ mesh with colors from red to blue denoting 
    the differences.
    Note that the abrupt changes (e.g., red to blue) are related to band crossings.
    c) $k$-dependent exciton wave function of the first bright (A) exciton
    on a logarithmic scale.
    Note that due to the magnetic field the $\pm$K degeneracy is lifted.
    The resulting exciton $g$ factor is calculated by
    multiplying the magnetic moments with the exciton wave functions (see Eq.~(\ref{eq:g})).
  }\label{fig:MoSe2_mk}
\end{figure*}
\section{Exciton $g$ factors of $\text{MoSe}_2$}
Using Eq.~(\ref{eq:g}) we are now able to calculate the
energy splitting of excitons in a magnetic field, i.e. its $g$ factors.
In Fig.~\ref{fig:MoSe2_mk}a,b) the resulting differences of the magnetic moments
between the valence band and the first and second conduction bands are shown.
These differences are weighted in Eq.~(\ref{eq:g})
by the square of the exciton wave function,
which is shown on a logarithmic scale in Fig.~\ref{fig:MoSe2_mk}c)
for the case of the exciton at K.
In Mo-based TMDCs the exciton transition to the lowest
conduction band is \textit{bright} due to the same spin character \cite{wang_-plane_2017,DarkExTr}, i.e. the so-called A exciton.
We find that the interband transition exactly at K
involves a change of the magnetic moment of $-2.3\,\mu_\text{B}$
while at $-$K the moment is reversed,
which would yield $g^\text{A}_\text{band} = 2 \cdot (-2.3) = -4.6$.
Away from $\pm$K the absolute value of the magnetic moment decreases
(see Supplement for details)
and if we take the ${\bf k}$ space dependent structure into account
we find a distinctly weaker $g^\text{A}$ factor of $-3.2$.
The second excited Rydberg state is considerably more extended in real
space and hence more localized in reciprocal space.
Consequently $g^{\text{A}^{2s}} = -3.7$ is much stronger,
even though stemming from the same bands \cite{Chen_2019}.
In the following we refer our wording to the absolute values of the $g$ factor.
The transition to the second conduction band (VB+1)
is the first \textit{dark} transition.
In Fig.~\ref{fig:MoSe2_mk}b) the calculated magnetic moments are shown.
Here, we find a distinctly larger difference of the magnetic moments of $-4.6\,\mu_\text{B}$ at K
and a decrease close to K as discussed before.
The weighted sum amounts to $g^\text{D} = -7.4$
(compared to $g^\text{D}_\text{band} = 2 \cdot (-4.6) = -9.2$).
Besides momentum direct excitons discussed before,
we can also use Eq.~(\ref{eq:g}) to evaluate indirect excitons \cite{exq}.
We find, e.g., for the lowest energy transitions $g = 12.0$ ($g_\text{band} = 15.6$)
for  K$\to$K' 
and $g = -6.0$ ($g_\text{band} = -8.6$) 
for K$\to\Lambda$.

\section{Comparison of different TMDCs}
Tab.~\ref{tab:mg} compiles our data of all five TMDC monolayers studied here.
We first focus on the difference of the transition exactly at K.
Changing the chalcogen atom from S to Se and further to Te
leads to an increase in $g^\text{A}_\text{band}$.
E.g. for Mo the value changes from $-4.3$, to $-4.6$ and $-4.9$, respectively.
If the metal atom is tungsten, the magnetic moments are smaller
and we find $-3.5$ and $-3.9$.
The transition from the second highest valence band (VB-1) at K (corresponding to the B exciton),
results in a very similar trend.
However, compared to $g^\text{A}_\text{band}$ the strength of the magnetic moment
is increased by $0.1$ to $0.4$.

\begin{table*}[tb]%
\centering
\caption{
Magnetic moments from Eq.~(\ref{eq:morb-new}) (in $\mu_\text{B}$) at the K point,
their differences $g^{\text{``A/B''}}_\text{band}$ which resemble the main contribution of the bright A and B transitions
(e.g. for MoS$_2$ $g^{\text{``A''}}_\text{band} = 2(m_\text{CB,K}-m_\text{VB,K})$ and $g^{\text{``B''}}_\text{band}= 2(m_\text{CB+1,K}-m_{\text{VB}-1\text{,K}})$),
and resulting $g$ factors from Eq.~(\ref{eq:g}).
In comparison several experimental measurements of the
$g$ factor of the A and B exciton are listed.
Note that for our magnetic moments $m$ we expect an error of about $\pm 0.05\,\mu_\text{B}$,
as well as about $\pm 0.1$ for the $g$ factors.
}\label{tab:mg}
  \begin{ruledtabular}
  \begin{scriptsize}
  \begin{tabular}{lcccccc}
    Material & $m_{\text{VB-1/VB},\text{K}}$ & $m_{\text{CB/CB+1},\text{K}}$ & $g^{\text{``A/B''}}_\text{band}$ & $g^{\text{A/B}}$ & Exp. $g^\text{A}$ & Exp. $g^\text{B}$\\
\hline
MoS$_2$ &$2.94$/$5.10$ & $2.98$/$0.76$ & $-4.24$/$-4.36$ & $-3.06$/$-3.10$ & $-1.7^\text{\cite{cadiz_excitonic_2017}}$,$-3.0^\text{\cite{Goryca_2019}}$,$-3.8^\text{\cite{Goryca_2019}}$,$-4.0^\text{\cite{stier_exciton_2016}}$,$-4.2^\text{\cite{Koperski_2018}}$,$-4.6^\text{\cite{PhysRevB.93.165412}}$ & $-4.3^\text{\cite{PhysRevB.93.165412}}$,$-4.65^\text{\cite{stier_exciton_2016}}$ \\
MoSe$_2$&$2.81$/$5.03$ & $2.74$/$0.46$ & $-4.58$/$-4.70$ & $-3.22$/$-3.28$ & $-3.8^\text{\cite{MacNeill_2015,Wang-mag_2015}}$,$-4.1^\text{\cite{Li_2014}}$,$-4.2^\text{\cite{Koperski_2018}}$,$-4.3^\text{\cite{Goryca_2019}}$,$-4.4^\text{\cite{PhysRevB.93.165412}}$ & $-4.2^\text{\cite{Koperski_2018}}$ \\
MoTe$_2$&$2.71$/$5.03$ & $2.60$/$0.21$ & $-4.86$/$-5.00$ & $-3.36$/$-3.36$ & $-4.6^\text{\cite{Arora_2016,Goryca_2019}}$ & $-3.8^\text{\cite{Arora_2016}}$ \\
WS$_2$  &$3.29$/$5.94$ & $1.35$/$4.21$ & $-3.46$/$-3.88$ & $-2.76$/$-2.80$ & $-3.7^\text{\cite{Koperski_2018}}$,$-3.94^\text{\cite{stier_exciton_2016}}$,$-4.0^\text{\cite{Goryca_2019}}$,$-4.25^\text{\cite{Plechinger_2016}}$,$-4.35^\text{\cite{Zipfel_2018}}$ & $-3.99^\text{\cite{stier_exciton_2016}}$,$-4.9^\text{\cite{Koperski_2018}}$\\
WSe$_2$ &$3.15$/$5.91$ & $0.99$/$3.97$ & $-3.88$/$-4.32$ & $-3.00$/$-3.22$ & $-1.6..-2.9^\text{\cite{Aivazian_2015}}$,$-3.2^\text{\cite{koperski2015single}}$,$-3.7^\text{\cite{Wang-mag_2015}}$,$-3.8^\text{\cite{Koperski_2018}}$,$-4.37^\text{\cite{Srivastava_2015}}$ & $-3.9^\text{\cite{Koperski_2018}}$ \\
  \end{tabular}
  \end{scriptsize}
  \end{ruledtabular}
\end{table*}%

As discussed above, the exciton $g$ factors are also sensitive to the region around K.
In all cases we find that the difference of the magnetic moments decreases away from K and
thus the $g$ factors are clearly smaller compared to the interband values $g_\text{band}$ at the K point.
For the Mo-based TMDCs we find $g^\text{A}$ ranges
between $-3.1$ and $-3.4$,
while WS$_2$ and WSe$_2$ have values of $-2.8$ and $-3.0$, respectively.
For the different materials the trends of the $g$ factors follow the trends described
for $g_\text{band}$ for both the A and B exciton.
The reduction of the exciton $g$ factor compared to $g_\text{band}$
is slightly stronger in Mo-based TMDCs and is approximately 30\%.

We note in passing that employing quasi-particle energies in Eq.~(\ref{eq:morb-new})
leads to $g$ factors that are slightly larger by about $0.2$,
i.e. $g^\text{A} = -3.2$, $-3.4$, $-3.6$, $-3.0$, and $-3.3$, respectively, for the materials listed in Tab.~\ref{tab:mg}.

Several experimental studies on exciton $g$ factors have been performed
in TMDC monolayers and their results are listed in Tab.~\ref{tab:mg}.
In general the data is quite scattered
and values between $-1.6$ and $-4.6$ are observed for the A excitons.
E.g. for MoSe$_2$ six measurements are available which range
between $-3.8$ and $-4.4$.
All in all, a one to one comparison of our results to experiment is not easily possible.
Within measurements from the same groups (e.g. \cite{Koperski_2018,Goryca_2019}) one can observe a weak
tendency that W-based TMDC have smaller values compared to Mo-based.
If we compare $g_\text{band}$ to experiment, one seems to find a very reasonable agreement.
The decrease of about 30\% of the exciton $g$ factors results
in generally slightly smaller values compared to the experiment.
However, we note that experiments are not performed for free-standing monolayers.
Additional dielectric screening (e.g. of the substrate)
results in a weakening of the exciton binding energy,
a larger spatial extent of the exciton,
a smaller extent in ${\bf k}$ space \cite{MoS2Dru},
and eventually in larger exciton $g$ factors.
For hBN substrate/encapsulation, e.g., the $g$ factor increases by about $0.1$/$0.2$.
We believe that the dependence of the $g$ factor
on the environment might partially explain the scattering of experiments in Tab.~\ref{tab:mg}.%

Furthermore, $g$ factors for higher excited Rydberg excitons ($2s$ etc.) have been measured
\cite{Chen_2019,Goryca_2019,Liu_Rydberg_2019}.
In most cases the $g$ factors of $2s$ excitons increase compared to $1s$.
As the spatial extent of these $ns$ excitons increases with $n$,
i.e. their extent in ${\bf k}$ space decreases \cite{qiu_screening_2016}.
This is perfectly in line with our results discussed above.

\section{Conclusion}
In summary, we have proposed an approach to calculate
magnetic moments and exciton $g$ factors of semiconductors from first principles.
Excluding excitonic effects,
we obtain $g_\text{band}$ factors ranging between $-3.5$ to $-4.9$ for monolayer WS$_2$ to MoTe$_2$, respectively.
Employing $GW$+BSE calculations
we find a distinct reduction of about 30\% resulting in $g$ factors 
which range between $-2.8$ and $-3.4$ for the excitons.
Compared to the experimental results,
our calculated values and trends are in good agreement
and open a pathway for better understanding
the change of optical properties of semiconductors in magnetic fields.

We note that calculations of $g_\text{band}$ for TMDCs
have recently been posted \cite{2002.02542,2002.11646,2002.11993}.

\section*{Acknowledgments}
We thank Ashish Arora for many constructive discussions and helpful comments.
The authors gratefully acknowledge the financial support from German Research Foundation (DFG project no. DE 2749/2-1),
the Collaborative Research Center SFB 1083 (project A13),
and the funding of computing time provided by the Paderborn Center for Parallel Computing (PC2).

\clearpage
\begin{center}\bf\large%
Supplement to\\
  Ab-initio studies of exciton $\texorpdfstring{\boldsymbol{g}}{g}$ factors:\\
  Monolayer transition metal dichalcogenides in magnetic fields
\end{center}
\setcounter{equation}{0}
\setcounter{figure}{0}
\setcounter{table}{0}
\setcounter{section}{0}
\makeatletter
\renewcommand{\thesection}{S\Roman{section}}
\renewcommand{\theequation}{S\arabic{equation}}
\renewcommand{\thefigure}{S\arabic{figure}}

\section{DFT and $\texorpdfstring{\boldsymbol{GW}\!}{GW}$/BSE calculations}
As a basis for the many-body calculations of the transition metal dichalcogenides (TMDCs)
we first carry out a DFT calculation in the
local density approximation (in the parametrization of Perdew and Zunger \cite{lda_pz}).
Norm-conserving pseudopotentials \cite{pp_ham} in the Kleinman-Bylander form \cite{pp_kb} are used.
The spin-orbit interaction is included in $j$-dependent pseudopotentials \cite{pp_so,so_Staerk_2011}
and is thus fully taken into account.
We employ a basis of 3 shells of Gaussian orbitals (30 functions 
with $s$, $p$, $d$, and $s^\ast$ symmetry for each atom, with decay constants 
from 0.16 to 2.5 $a_\text{B}^{-2}$).
In our approach, the wave function of the crystal is given by \cite{wieferink_improved_2006}
\begin{align*}
  \Psi_{n{\bf k}}({\bf r}) = \sum_{\alpha\mu} c_{\alpha\mu}^{n,{\bf k}} \chi^{\bf k}_{\alpha\mu}({\bf r}) ,
\end{align*}
where $\alpha$ denotes the orbital and ${\boldsymbol \tau}_\mu$ the position of the basis atom.
$n$ and ${\bf k}$ are the band number and the ${\bf k}$ point,
while $c_{\alpha\mu}^{n,{\bf k}}$ are the coefficients resulting from the variation of the trial functions
containing Gaussian orbitals $\varphi_{\alpha\mu}$
\begin{align*}
  \chi^{\bf k}_{\alpha\mu}({\bf r}) = \sum_{{\bf R}_j} e^{i{\bf k}({\boldsymbol \tau}_\mu + {\bf R}_j)} \varphi_{\alpha\mu}({\bf r} - {\boldsymbol \tau}_\mu - {\bf R}_j) .
\end{align*}
For the DFT calculations of TMDCs, a mesh of $12\times 12$ ${\bf k}$-points is employed
in the two-dimensional Brillouin zone.

For the following $GW\!$/BSE calculations we utilize the $GdW$ approach \cite{GdW}.
This has been successfully used to describe the electronic and optical properties
of TMDCs as discussed in detail in Ref.~\cite{GdWTMDC}.

\section{Calculation of the magnetic moments in the local approximation}
We evaluate the matrix elements of the angular momentum operator in $z$ as
\begin{widetext}
\vspace*{.1cm}
\begin{align}
  \langle \hat{L}_z \rangle^{\bf k}_{\alpha\mu,\alpha'\mu'} &= \int_\text{uc} d^3r\, \chi^{{\bf k}*}_{\alpha\mu}({\bf r}) \hat{L}_{z} \chi^{{\bf k}}_{\alpha'\mu'}({\bf r}) \nonumber\\
  &=       \sum_{{\bf R}_j{\bf R}_{j'}}   e^{i{\bf k}({\boldsymbol \tau}_{\mu'} + {\bf R}_{j'} - {\boldsymbol \tau}_\mu - {\bf R}_j)} \int_\text{uc} d^3r\, \varphi^*_{\alpha\mu}({\bf r} - {\boldsymbol \tau}_\mu - {\bf R}_j) \lbrace x\hat{p}_y - y\hat{p}_x \rbrace \varphi_{\alpha'\mu'}({\bf r} - {\boldsymbol \tau}_{\mu'} - {\bf R}_{j'}) \nonumber\\
  &\approx \sum_{{\bf R}_j \in \text{NN}} e^{i{\bf k}({\boldsymbol \tau}_{\mu'}                - {\boldsymbol \tau}_\mu + {\bf R}_j)} \int_{\mathbb{R}^3} d^3r\, \varphi^*_{\alpha\mu}({\bf r} - {\boldsymbol \tau}_\mu) \lbrace x\hat{p}_y - y\hat{p}_x \rbrace \varphi_{\alpha'\mu'}({\bf r} - {\boldsymbol \tau}_{\mu'} - {\bf R}_{j})
  =: \langle \hat{L}_z \rangle^{\text{loc,}\bf k}_{\alpha\mu,\alpha'\mu'} .
\end{align}
\vspace*{.1cm}
\end{widetext}
We note that the last step is an approximation
and the extra terms due to the transformations ${\bf r} \to {\bf r} - {\bf R}_j$ etc. are neglected
and only contributions from nearest neighbours are taken into account symmetrically.
When including these terms, the resulting sums are hard to converge
and the evaluation using derivatives by $k$ is more convenient
as described in the main text.

We note that even if the magnetic moments in this local approximation
have a large discrepancy, its differences and the excitonic $g$ factors
are in reasonable agreement to Eq.~(\ref{eq:morb-new}).
We find $g^\text{A}_\text{loc} = -3.5$, $-3.5$, $-3.3$, $-3.3$, and $-3.4$ (in the order of Tab.~\ref{tab:mg})
which are slightly larger due to less strongly varying magnetic moments close to $\pm$K.

\section{Convergence of the magnetic moments}
In our work we typically use a local Gaussian basis set with orbitals $l_\text{max}\leq 2$.
This allows a precise evaluation of the wave function
which is sufficient for the calculation of the local magnetic moments.
Considering the full Bloch state,
we introduce a complete orthogonal set $1 = \sum_{n'} |u_{n'{\bf k}} \rangle \langle u_{n'{\bf k}}|$
in the full evaluation of
\begin{align}\label{eq:morb-new-real}
m_{n{\bf k}}^\text{orb} = \mu_\text{B} & \sum_{n'} (E_{n'{\bf k}} - E_{n{\bf k}}) \times \nonumber\\
  &\text{Im} \left( \left\langle \frac{\partial u_{n{\bf k}}}{\partial k_x} \middle| u_{n'{\bf k}} \right\rangle \left\langle u_{n'{\bf k}} \middle| \frac{\partial u_{n{\bf k}}}{\partial k_y} \right\rangle \right) .
\end{align}
However, for the completeness assumed here,
we find that $l_\text{max}\leq 2$ is not sufficient
and we have implemented and employed $l_\text{max}\leq 4$, i.e. 35 functions with $s$, $p$, $d$, $f$, and $g$ symmetry.
Note that the difference quotient is given by the finite-difference \cite{Sai_2002}
for which the phase has to be aligned.

\begin{figure}[t]
  \centering
  \includegraphics[width=.5\linewidth]{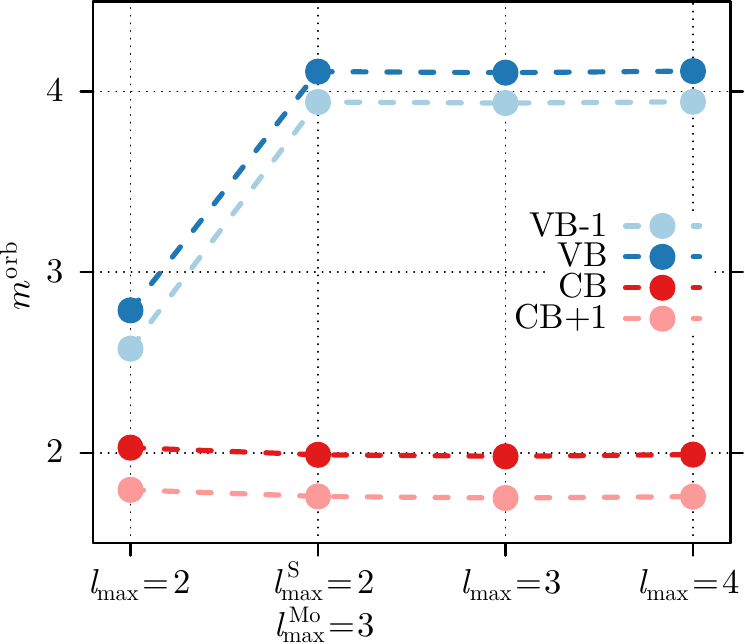}
  \caption{
    $m^\text{orb}$ at the K point of the four valance and conduction bands, respectively.
    The results for the MoS$_2$ monolayer for different $l_\text{max}$ of different Gaussian basis sets
    are connected by dashed lines as guide to the eye.
    Compared to Tab.~1 (main text) the spin part is still missing.
  }\label{fig:morb-conv}
\end{figure}

The resulting magnetic moments depending on $l_\text{max}$ are shown in Fig.~\ref{fig:morb-conv}.
While the usage of $f$ orbitals at the Mo atoms is essential for the valence bands,
the results are well converged including further orbitals.
We note that it is essential to include all bands $n'$ in Eq.~(\ref{eq:morb-new-real})
as the factor of the energy difference increases for the higher-lying bands.
When trying to use less bands (not shown),
we find that $m^\text{orb}$ of the valence bands decreases distinctly
and the resulting magnetic moments and $g$ factors are smaller compared to the converged results.

\begin{figure}[t]
  \centering
  \includegraphics[width=.5\linewidth]{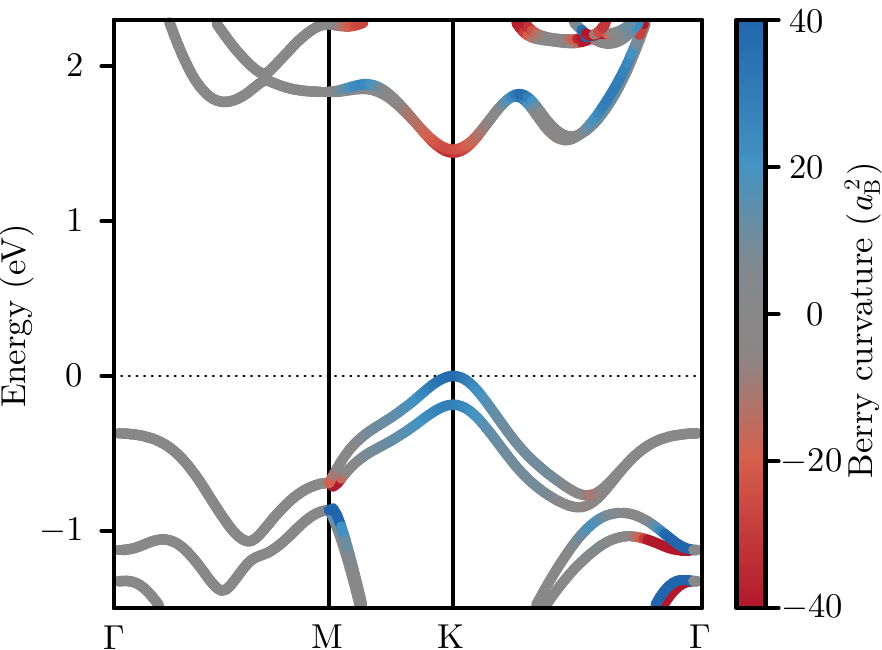}
  \caption{
    Berry curvature of MoSe$_2$ corresponding to Fig.~1 in the main text.
    The results are in good agreement with Feng et al. \cite{Feng_2012}.
    In regions with large Berry curvatures the local approximation
    is particularly bad.
  }\label{fig:MoSe2-berry}
\end{figure}

As addressed in the main text,
the magnetic moments especially at $\pm$K are effected by the Berry curvature.
The curvature for the monolayer MoSe$_2$ is shown in Fig.~\ref{fig:MoSe2-berry}.

The evaluation of Eqs.~(4) and (5)
is mostly carried out within DFT in the main text.
By employing many-body perturbation theory
in the $GW$ approximation \cite{HedinGW,GdWTMDC},
we can correct for the underestimated band gap.
For TMDC monolayers it has been found that
$\Psi^\text{GW} \approx \Psi^\text{DFT}$ is a good approximation,
which means that only the energies $E_{n{\bf k}}$ change.
In Tab.~\ref{tab:mgw} we present the resulting
magnetic moments and exciton $g$ factors.
We have employed a scissors shift
after we have checked that the result is almost
identical to band dependent energy shifts.

The magnetic moments of the lowest conduction band have been measured indirectly.
Values of $1.84$, $1.08$, and $1.3\,\mu_\text{B}$ have been reported for
MoSe$_2$, WS$_2$, and WSe$_2$ \cite{Koperski_2018,Lyons_2019}, respectively.
We calculate magnetic moments of $2.7$, $1.4$, and $1.0\,\mu_\text{B}$, respectively,
when Bloch states are taken into account (Tab.~1, main text).
We conclude that we find reasonable agreement
in contrast to the local approximation in which we observe approximately $+1$, $-1$, and $-1\,\mu_\text{B}$, respectively.
Employing the $GW$ approximation (Tab.~\ref{tab:mgw}) the magnetic moments are clearly larger.

\begin{table}[tb]%
\centering
\caption{
Magnetic moments from Eq.~(4) (in $\mu_\text{B}$) at the $K$ point,
their differences $g^{\text{``A/B''}}_\text{band}$,
and resulting $g$ factors from Eq.~(7)
employing the $GW$ approximation.
See Tab.~I of the main text.
}\label{tab:mgw}
  \begin{ruledtabular}
  \begin{tabular}{lcccc}
    Material & $m_{\text{VB-1/VB},\text{K}}$ & $m_{\text{CB/CB+1},\text{K}}$ & $g^{\text{``A/B''}}_\text{band}$ & $g^{\text{A/B}}$\\
\hline
MoS$_2$ &$4.0$/$6.4$ & $4.2$/$1.8$ & $-4.4$/$-4.4$ & $-3.2$/$-3.3$\\
MoSe$_2$&$4.0$/$6.6$ & $4.2$/$1.6$ & $-4.8$/$-4.8$ & $-3.4$/$-3.5$\\
MoTe$_2$&$4.0$/$6.9$ & $4.4$/$1.4$ & $-5.0$/$-5.2$ & $-3.6$/$-3.6$\\
WS$_2$  &$4.8$/$8.4$ & $2.7$/$6.5$ & $-3.8$/$-4.2$ & $-3.0$/$-3.0$\\
WSe$_2$ &$4.6$/$8.6$ & $2.3$/$6.5$ & $-4.2$/$-4.6$ & $-3.3$/$-3.5$\\
  \end{tabular}
  \end{ruledtabular}
\end{table}%

\section{Convergence of the exciton $g$ factors}
Employing the Bethe-Salpeter equation
(which includes the screened direct and the bare exchange interaction)
the excitons are described using a $N \times N$ mesh,
i.e. $N\times N$ cells in real space are evaluated
and $N\times N$ points raster the Brillouin zone in reciprocal space (see Fig.~2c in the main text,
we solve the BSE in the entire Brillouin zone).
The differences of the magnetic moments $\Delta m$ and thus the resulting exciton $g$ factors
depend on the chosen ${\bf k}$ points.

\begin{figure}[t]
  \centering
  \includegraphics[width=.5\linewidth]{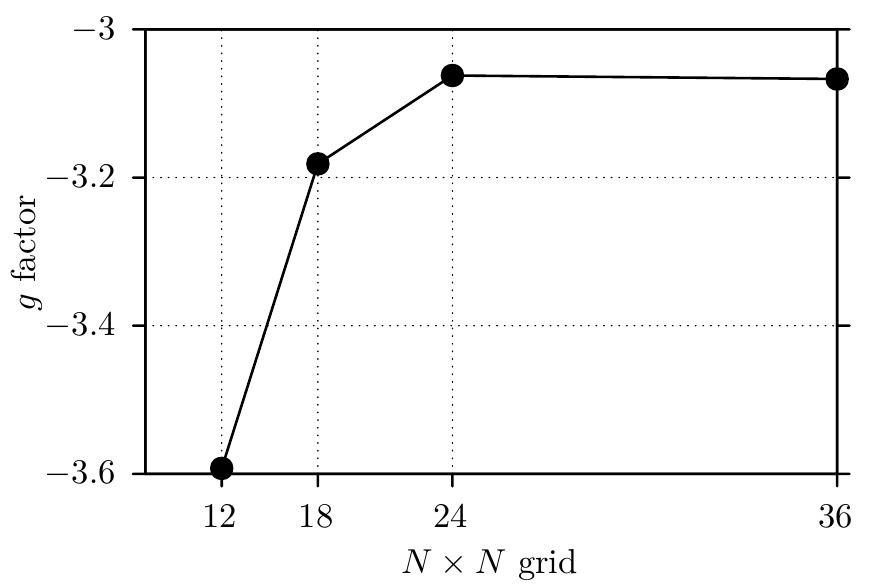}
  \caption{
    $g$ factor of the A exciton of the MoS$_2$ monolayer for different $N\times N$ grids.
  }\label{fig:g-conv}
\end{figure}

In Fig.~\ref{fig:g-conv} we compare the resulting $g$ factor for different grids.
We find that our result is converged to better than $0.02\,\mu_\text{B}$
for the employed $24\times 24$ mesh in Fig.~2 (main text)
which we employ for all materials in the main text.
\\[-.6cm]

\section{Contributions of the magnetic moments}
The magnetic moment consists of a spin part $m^\text{spin}$
and an orbital part $m^\text{orb}$.
The latter one (as resulting from
Eq.~(5), main text) can be understood as 
a local orbital part $m^\text{orb}_\text{loc}$
(Eq.~(3) in main text)
and a remaining correction arising due to
the different contributions of $\hat{L}_z$ in different unit cells.
The second contribution
$m^\text{val} := m^\text{orb} - m^\text{orb}_\text{loc}$
corresponds to the \textit{valley} term discussed in Ref. \cite{Koperski_2018}.

\begin{figure*}[h]
  \centering
  \includegraphics[width=.95\linewidth]{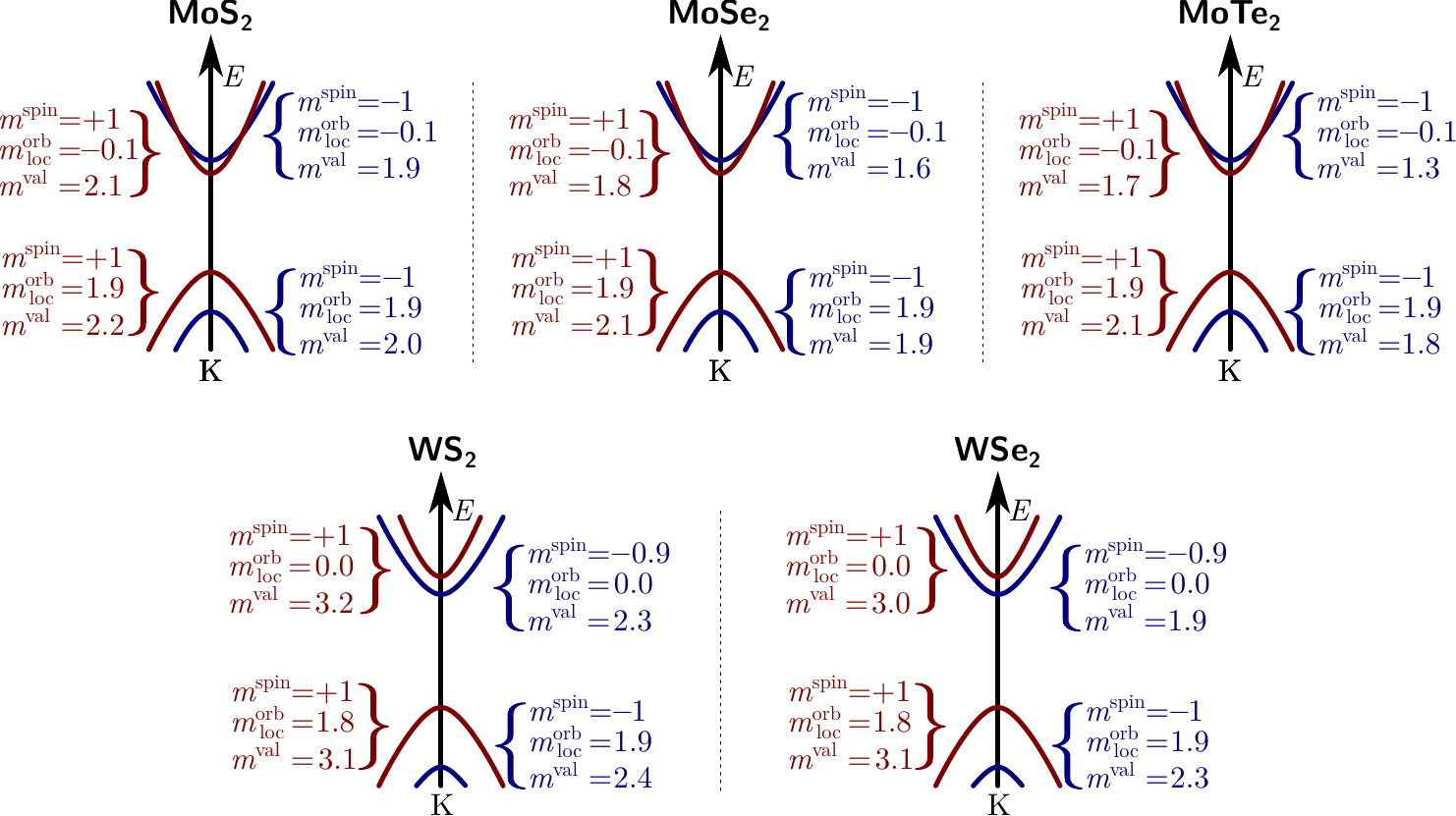}
  \caption{
    The different contributions of the magnetic moments $m^\text{spin}$, $m^\text{orb}_\text{loc}$ (Eq.~(3), main text),
    and $m^\text{val} = m^\text{orb} - m^\text{orb}_\text{loc}$ (Eq.~(5), main text) are
    given for each band at the K point (in $\mu_\text{B}$).
    The sketched bands are colored according to the spin character.
    The sums $m = m^\text{orb} + m^\text{spin}$ are given in Tab.~1 of the main text.
    Note that excitonic effects are missing.
  }\label{fig:TMDC_m_cont}
\end{figure*}

\begin{figure*}[t]
  \centering
  \includegraphics[width=.66\linewidth]{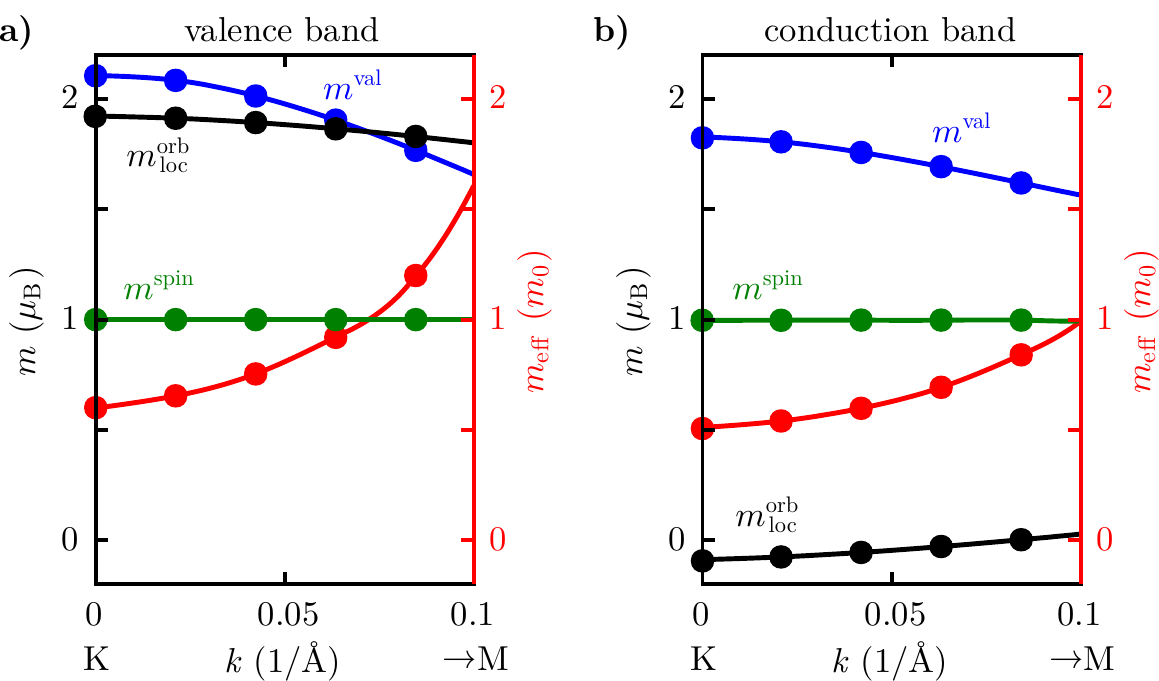}
  \caption{
    The different contributions of the magnetic moments $m^\text{spin}$, $m^\text{orb}_\text{loc}$ (Eq.~(3), main text),
    and $m^\text{val} = m^\text{orb} - m^\text{orb}_\text{loc}$ (Eq.~(5), main text)
    are plotted in the KM direction close to the K point.
    In addition the effective masses of the bands are shown with its scale on the right side.
    In a) and b) the highest valence band and lowest conduction band of MoSe$_2$ are shown.
  }\label{fig:MoSe2_m-detail}
\end{figure*}

In Fig.~\ref{fig:TMDC_m_cont} we present
the contributions for all five TMDCs
while in Fig.~\ref{fig:MoSe2_m-detail}
we show the $k$ dependence of these contributions.
We note that $m^\text{val}$ and the effective mass
show opposite trends
with a stronger gradient of the effective mass.

\clearpage

\end{document}